\documentclass[a4paper,11pt]{article}
\usepackage{pos}
\usepackage{aas_macros}

\title{Exploring the Galactic Plane: A Comparative Study of \textit{Fermi}-LAT Sources and H.E.S.S.'s Non-Detection at TeV Energies}
\ShortTitle{4FGL and HGPS ULs comparison}

\author*[a]{François Brun}
\author[b]{Baptiste Le Nagat-Neher}
\author[c]{Marianne Lemoine-Goumard}%28
\author[c]{Marie-Hélène Grondin}%28
\author[c]{Paul Fauverge}%28

\affiliation[a]{IRFU, CEA, Université Paris-Saclay, 
                F-91191 Gif-sur-Yvette, France}

\affiliation[b]{LUX, Observatoire de Paris, Université PSL, Sorbonne Université, CNRS, 92190
Meudon, France}

\affiliation[c]{Université Bordeaux, CNRS, LP2I Bordeaux, UMR 5797,
                33170 Gradignan, France}

\emailAdd{francois.brun@cea.fr}

\abstract{The HESS Galactic Plane Survey (HGPS), published in 2018, presented a decade of very-high-energy (VHE) gamma-ray observations along the Galactic plane. This study was accompanied by the release of several maps in FITS format, offering a detailed view of the region. The flux upper-limits from these HGPS maps can be compared to the high-energy (HE) spectra of sources catalogued by the \textit{Fermi}-LAT in the same region. For some sources, extrapolating the \textit{Fermi}-LAT flux into the VHE range predicts flux values exceeding the upper-limits set by HESS. In this work, we present the results of this comparison and highlight the sources that are of particular interest for future VHE observations.}

\ConferenceLogo{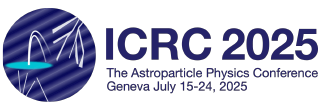}

\FullConference{39th International Cosmic Ray Conference (ICRC2025)\\
 15–24 July 2025\\
Geneva, Switzerland\\}
\begin{document}
\maketitle

\section{Introduction}

The H.E.S.S. Galactic Plane Survey \cite{HGPS_Paper} is a deep survey of the inner Milky Way conducted using the High Energy Stereoscopic System (H.E.S.S.), a ground-based array of Cherenkov telescopes in Namibia. It maps very-high-energy (VHE; E$> 100$GeV) gamma rays across the Galactic plane. The HGPS provides a catalog of gamma-ray sources and was accompanied by the release of several maps\footnote{\url{https://www.mpi-hd.mpg.de/HESS/hgps/}} such as significance maps of the signal, flux and flux upper limit (UL) maps. The region covered by this study was ranging from $l = 246^\circ$ to $75^\circ$ and $|b| < 5^\circ$, in  Galactic coordinates. The maps have a binning of $0.02^\circ$ / pixel.

The Fourth Fermi Large Area Telescope Source Catalog – Data Release 4 (4FGL-DR4) is the latest and most comprehensive catalog of gamma-ray sources detected by the Fermi Gamma-ray Space Telescope \cite{Fermi-DR4}. Initially released in 2023, it is based on 14 years of all-sky observations by the Large Area Telescope (LAT), covering the energy range from 50 MeV to 1 TeV. In its latest revision\footnote{\url{https://fermi.gsfc.nasa.gov/ssc/data/access/lat/14yr_catalog/}}, the 4FGL-DR4 catalog includes a total of 7,194 sources.
Each source in the catalog is characterized by detailed spectral and temporal information, including photon fluxes, energy spectra, variability indices, and positional uncertainties. The 4FGL-DR4 uses the Pass 8 event-level analysis \cite{Pass8}, which provides enhanced sensitivity and improved energy and spatial resolution compared to earlier data reconstructions.

The aim of the study is to compare the spectra of the \textit{Fermi}-LAT detected sources with the flux ULs set by H.E.S.S. in the Galactic Plane Survey. 

\section{Analysis}

\subsection{Source selection}

The first step for the comparison is to define the spatial regions in which it can be done. From the HGPS significance map derived with a correlation radius of $R_c = 0.1^\circ$, an exclusion mask was defined from the pixels with values above $5\sigma$. This exclusion mask was then expanded by $0.3^\circ$ to follow the procedure described in the HGPS analysis (see Sect. 3.2.2 of \cite{HGPS_Paper}). In addition, we considered only the pixels more than $0.1^\circ$ away from a pixel with null significance, and containing a flux UL value different from zero. 

\begin{figure*}%[t!]
\centering
        \includegraphics[width=0.99\textwidth]{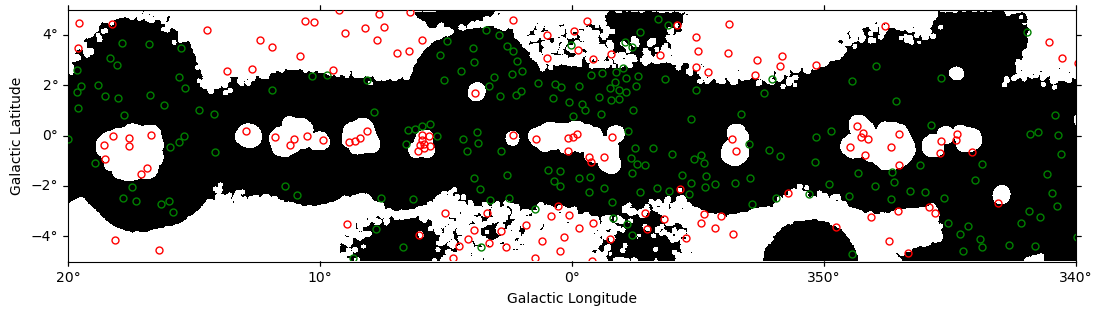}
    \caption{The map shows the region around the Galactic Center. The black regions corresponds to regions where no source is detected in the HGPS, where the HESS significance is different from 0 and where the HGPS flux UL value is different from 0. The green points indicate the 4FGL sources that are selected for the HGPS-4FGL spectral comparison, while the red one are those excluded for this comparison.}
    \label{fig_sourcepos}
\end{figure*}

Then, the 4FGL sources selected for the comparison were the point-like sources (without a ``c'' or ``e'' flag) inside the accepted region. The 4FGL sources with a best-fit spectral model including a cutoff were excluded as they are not expected to show significant emission in the VHE range.

This procedure led to a total of 418 selected sources. Figure \ref{fig_sourcepos} shows the positions of the 4FGL sources in the region of the HGPS maps around the Galactic Center.

\subsection{Spectra comparison}

From the HGPS, the flux UL map gives access to the $95\%$ Confidence Level UL on the integral flux above 1 TeV, under the assumption of a power-law spectrum with a spectral index of $2.3$. The maps were used to derive the UL ($95\%$ CL) on the differential flux at 1 and 10 TeV.
As stated in section $\mathrm{7.1}$ of the HGPS publication \cite{HGPS_Paper}, the flux-like values in the maps computed with a correlation radius of $0.1^\circ$ give only about $80\%$ of those of a point-like source, due to the H.E.S.S. Point Spread Function. The values extracted were therefore corrected by applying a factor of $1.25$.

From the 4FGL catalog used, the spectral parameters of each selected source were extracted. From the best-fit spectral model, the flux value at 1 TeV can be compared to the HGPS UL value. If the UL is below the lower bound of the 1-$\sigma$ confidence interval of the \textit{Fermi}-LAT measurement, the HGPS value is considered as constraining. Table \ref{table_sources} lists the 13 sources for which the HGPS measurement is constraining the spectrum provided in the 4FGL-DR4. 

\begin{table}[htbp]
\centering
\footnotesize
\begin{tabular}{l|l|l|l|l|l}
Source & GLON & GLAT & Signif. & HESS UL (1 TeV) & Association \\
 & [deg] & [deg] &  & [ph/cm$^{2}$/s/TeV] &  \\
\hline
4FGL J0928.4-5256 & 275.101 & -1.4034246 & 12.52 & $1.45\cdot 10^{-13}$ & -\\
4FGL J1115.1-6118 & 291.6333 & -0.5680487 & 16.89 & $1.43\cdot 10^{-13}$ & FGES J1109.4-6115 field \\
4FGL J1216.4-6317 & 299.0078 & -0.68271774 & 4.33 & $2.07\cdot 10^{-13}$ & -\\
4FGL J1524.8-5904 & 321.5396 & -1.854237 & 12.38 & $9.44\cdot 10^{-14}$ & PMN J1524-5903 \\
4FGL J1712.9-4105 & 346.18567 & -1.1619184 & 4.50 & $1.47\cdot 10^{-13}$ & -\\
4FGL J1719.0-4038 & 347.21805 & -1.8373653 & 4.21 & $4.20\cdot 10^{-14}$ & -\\
4FGL J1734.6-2912 & 358.49554 & 1.8873749 & 4.64 & $4.60\cdot 10^{-14}$ & -\\
4FGL J1735.9-3342 & 354.86386 & -0.77305466 & 5.61 & $6.28\cdot 10^{-14}$ & SNR G354.8-00.8 \\
4FGL J1741.8-2537 & 2.3833275 & 2.4545393 & 13.48 & $9.16\cdot 10^{-14}$ & NVSS J174154-253743 \\
4FGL J1744.5-2612 & 2.2042766 & 1.6288257 & 6.98 & $6.50\cdot 10^{-14}$ & 1RXS J174419.9-261230 \\
4FGL J1829.3-1614 & 15.9977 & -2.6058545 & 4.62 & $2.00\cdot 10^{-13}$ & SNR G016.2-02.7 \\
4FGL J1830.1-1440 & 17.47052 & -2.0428274 & 4.33 & $7.66\cdot 10^{-14}$ & -\\
4FGL J1830.3-1601 & 16.295761 & -2.7035134 & 4.12 & $1.35\cdot 10^{-13}$ & SNR G016.2-02.7 \\
\hline
\end{tabular}
\caption{Selected 4FGL sources with H.E.S.S. upper limits at 1 TeV and known associations. The Source name is the identifier from the 4FGL-DR4 catalog, the GLON and GLAT columns indicate the Galactic longitude and latitude of the source. The Significance column lists the sources average significance and the HESS UL indicates the flux UL on the differential flux at 1 TeV with an assumed spectral index of $2.3$, after application of a factor of 1.25 (see text). The last column lists the associated sources from the 4FGL-DR4 catalog.}
\label{table_sources}
\end{table}

\section{Discussion}

Most sources in the selected subset have moderate detection significances in the 4FGL-DR4 catalog (typically between 4 and 17 sigma, with a median value of 4.64). Several sources are spatially associated with known astrophysical objects, such as supernova remnants (e.g., SNR G354.8-00.8 and SNR G016.2-02.7), radio sources (e.g., PMN J1524-5903, NVSS J183540-054452), or X-ray sources (e.g., 1RXS J180938.8-244415). One source (Pal 6) is associated with a globular cluster.

%%%More info from points 
The 4FGL spectral points are also available in the data release. They are computed in pre-defined energy bins and the flux value as well as the flux errors and significance in the bin are provided for each bin. Based on the spectral points, it is possible to further assess whether a tension exists between the \textit{Fermi}-LAT and the H.E.S.S. measurements. The SEDs of the sources for which the HGPS measurement is constraining are presented 
in Fig. \ref{fig_specs1} and Fig. \ref{fig_specs2}. 

\begin{figure*}%[t!]
\centering
        \includegraphics[width=0.49\textwidth]{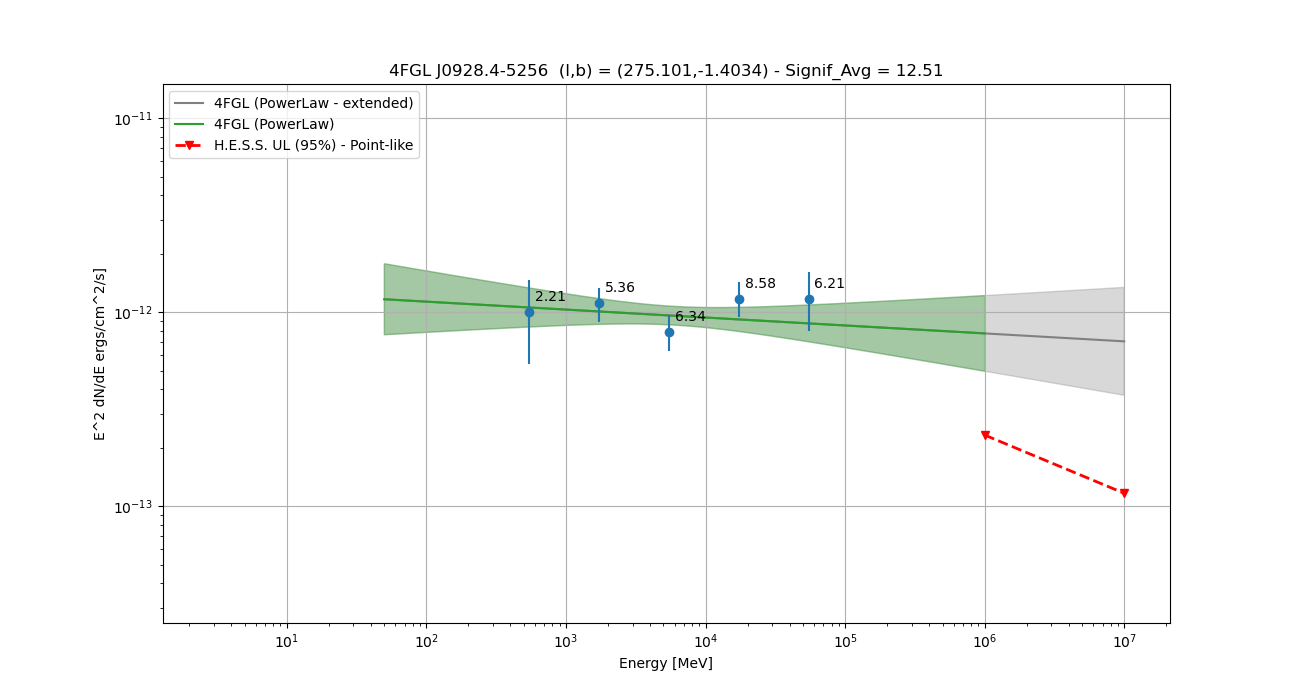}
        \includegraphics[width=0.49\textwidth]{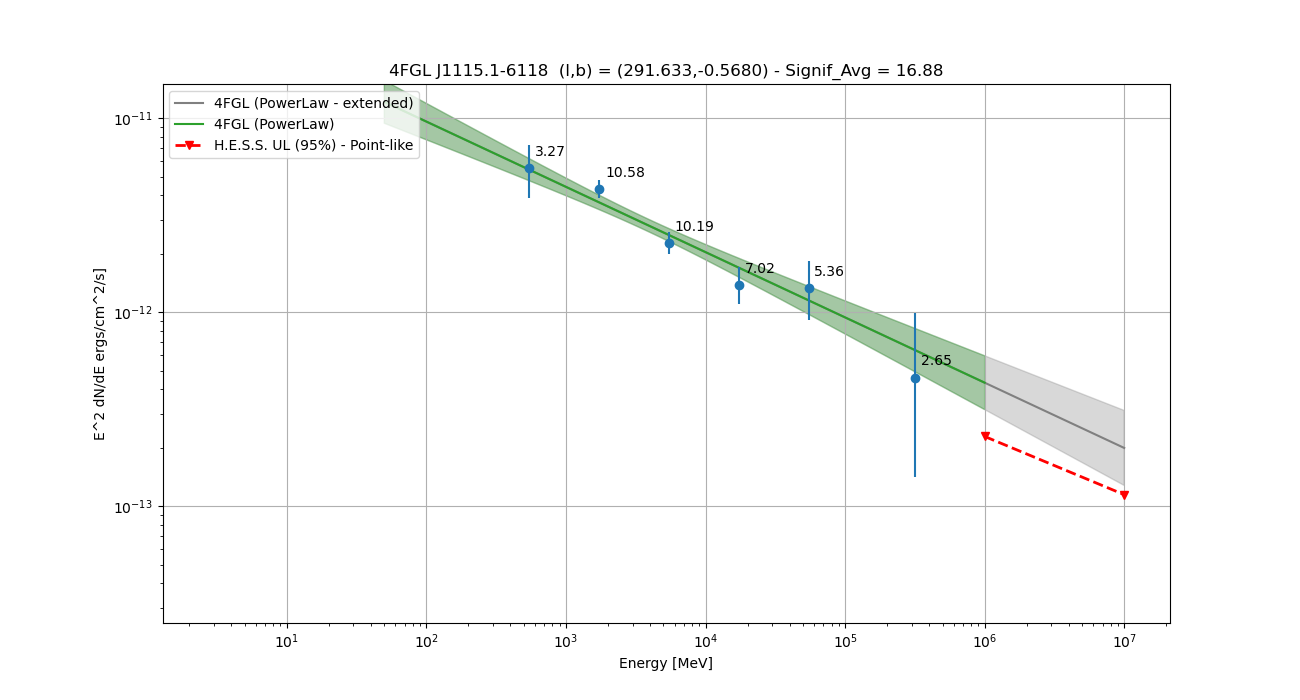}\\
        \includegraphics[width=0.49\textwidth]{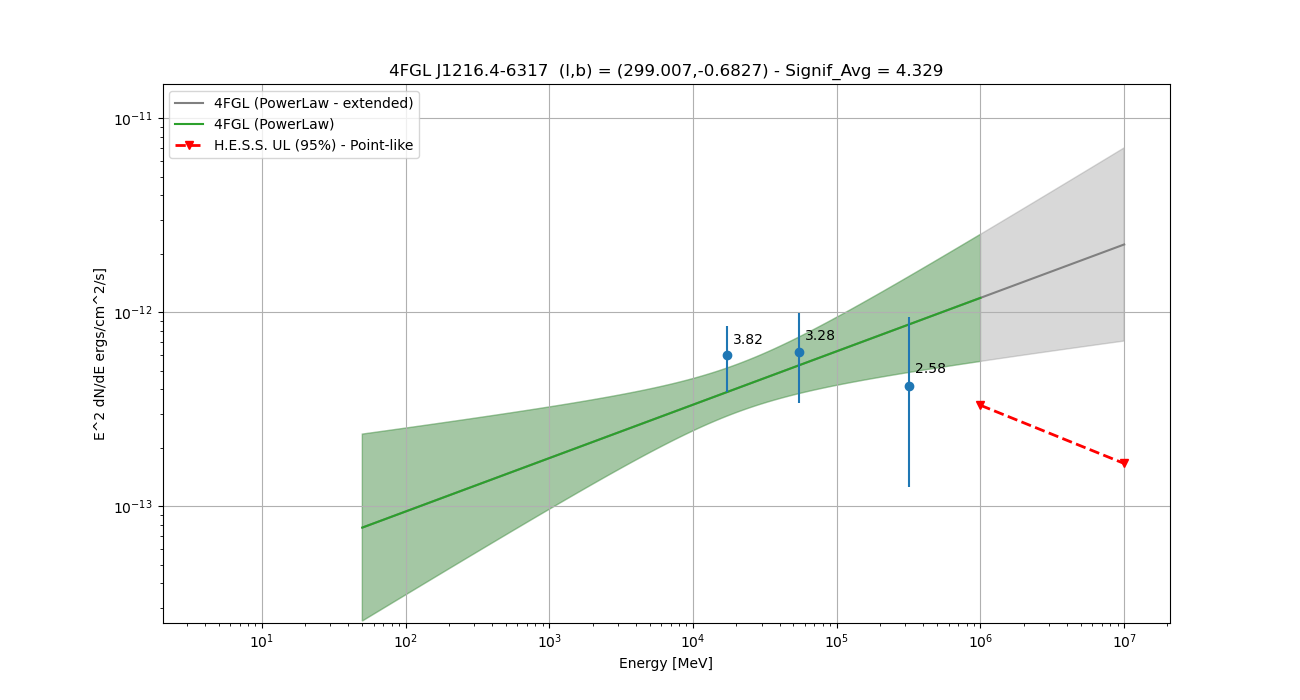}
        \includegraphics[width=0.49\textwidth]{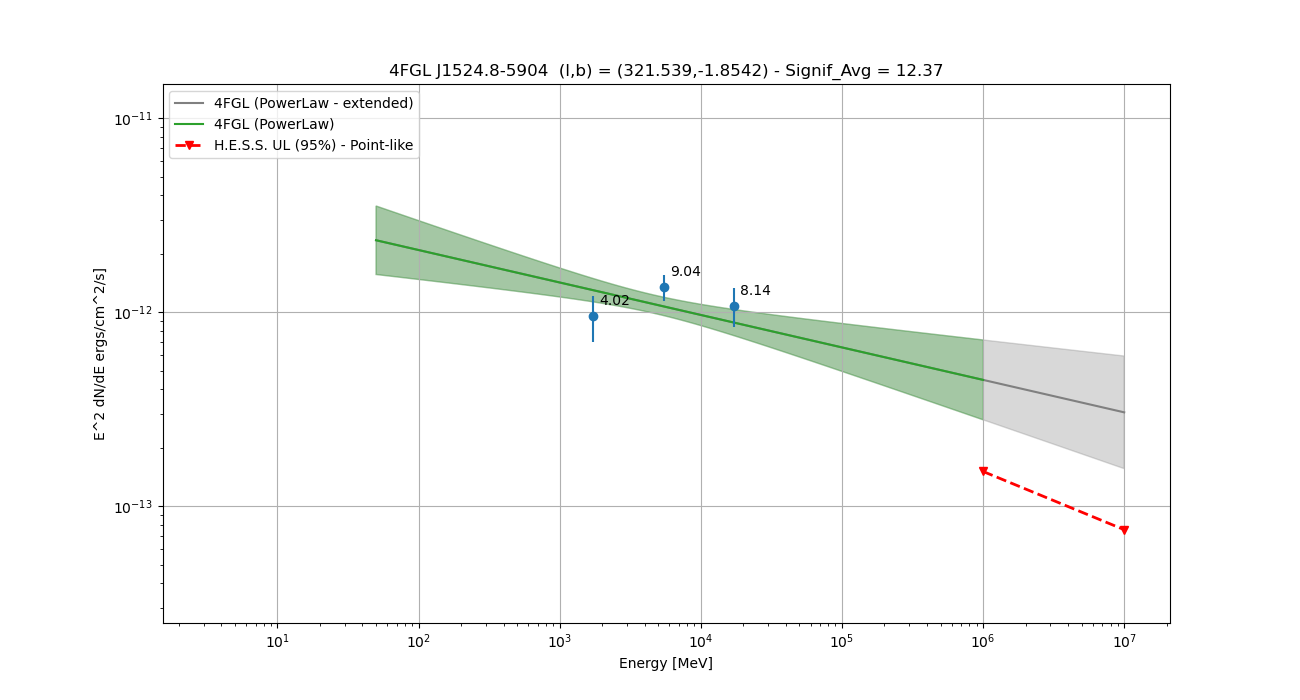}\\
        \includegraphics[width=0.49\textwidth]{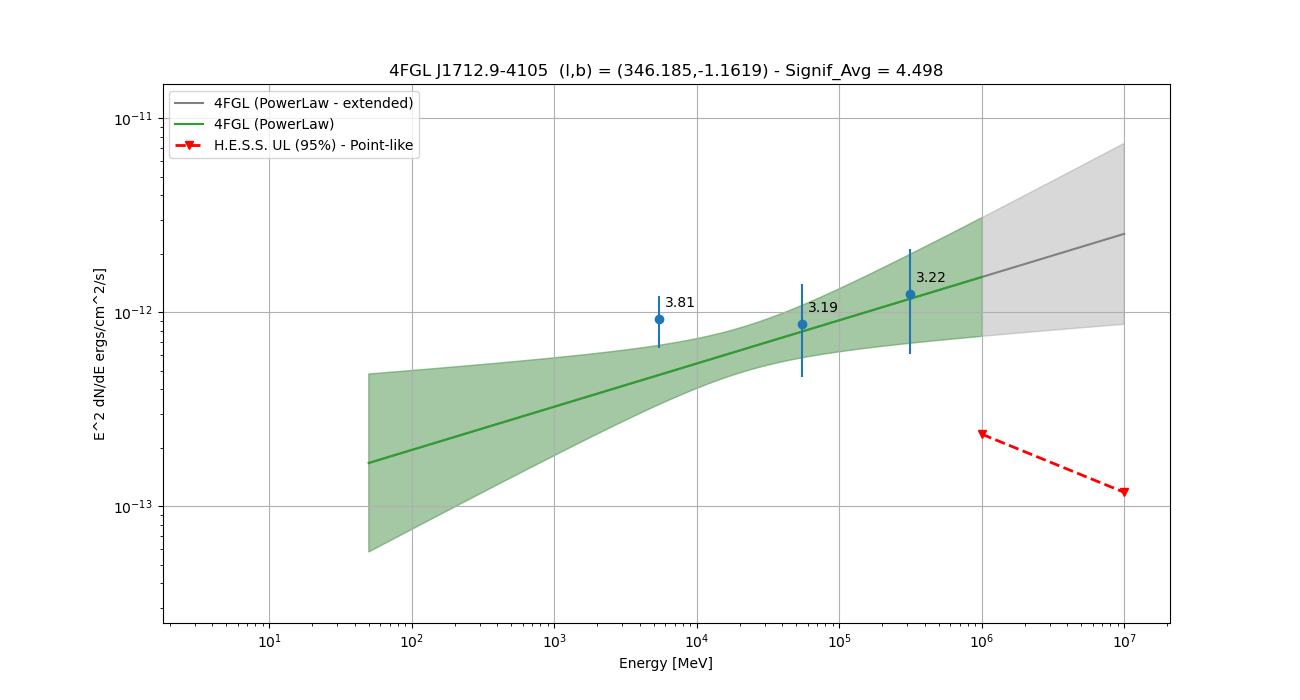}
        \includegraphics[width=0.49\textwidth]{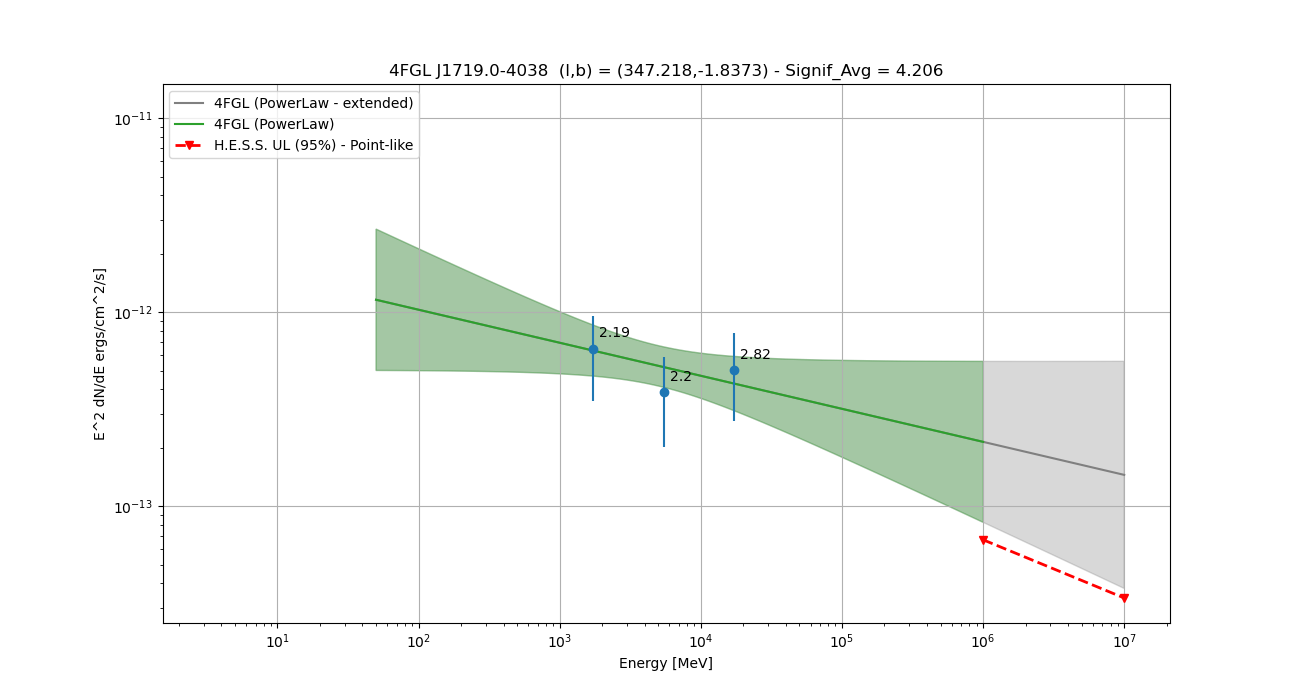}\\
        \includegraphics[width=0.49\textwidth]{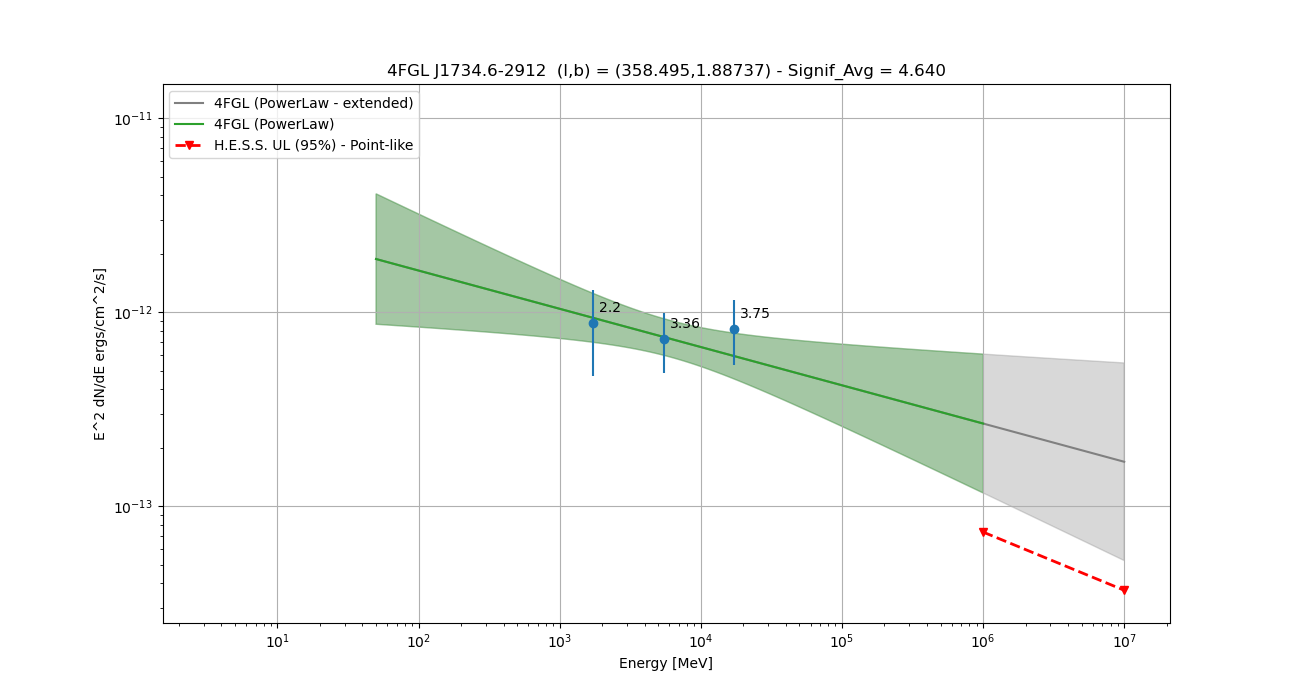}
        \includegraphics[width=0.49\textwidth]{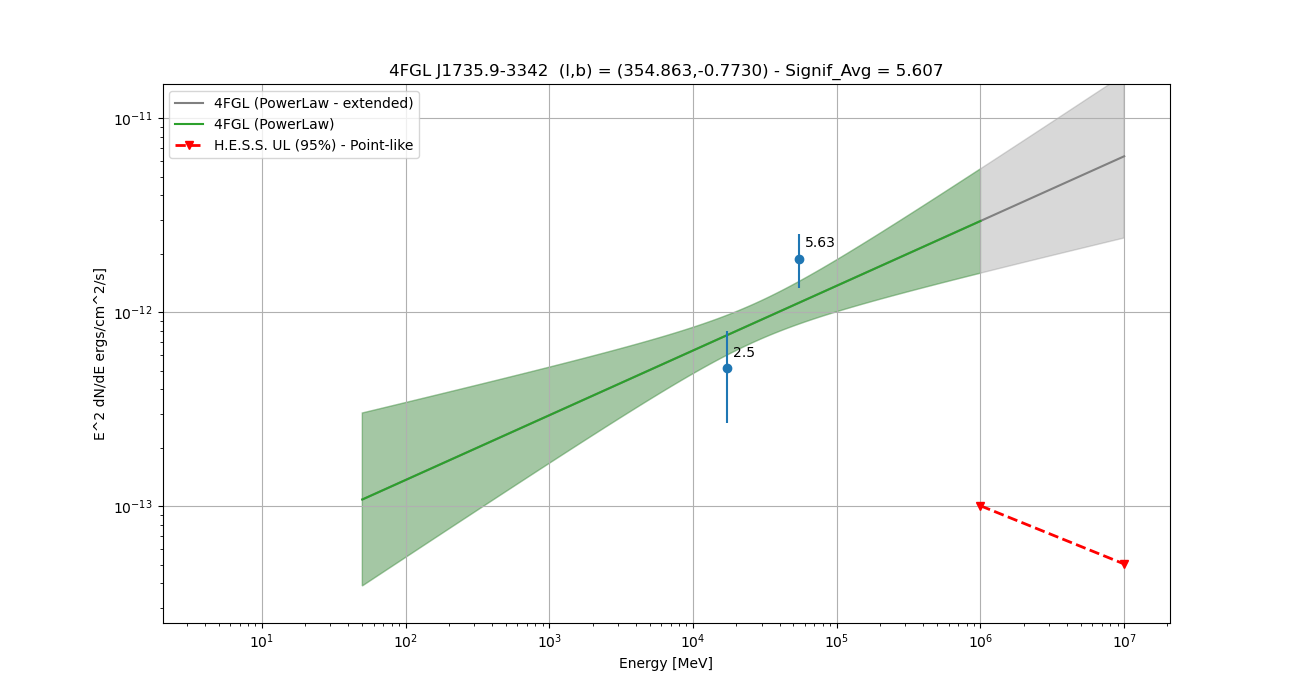}\\
    \caption{SED of Fermi-LAT sources identified as constrained by the HGPS data. The green butterfly displays the best-fit Power-Law spectrum from the \textit{Fermi}-LAT between $50$ MeV and $1$ TeV and the grey one shows an extrapolation of this measurement up to $10$~TeV. The red triangles show the H.E.S.S. flux ULs at 95\% CL at 1 and 10 TeV, scaled by a factor $1.25$. The blue points are the \textit{Fermi}-LAT spectral points derived in pre-defined energy bins. A point is displayed on the figure only if the \texttt{Sqrt\_TS\_band} parameter in the 4FGL-DR4 catalog is above 2. The value of this parameter -- which is indicative of the significance of the point -- is indicated next to the point.}
    \label{fig_specs1}
\end{figure*}

\begin{figure*}%[t!]
\centering
        \includegraphics[width=0.49\textwidth]{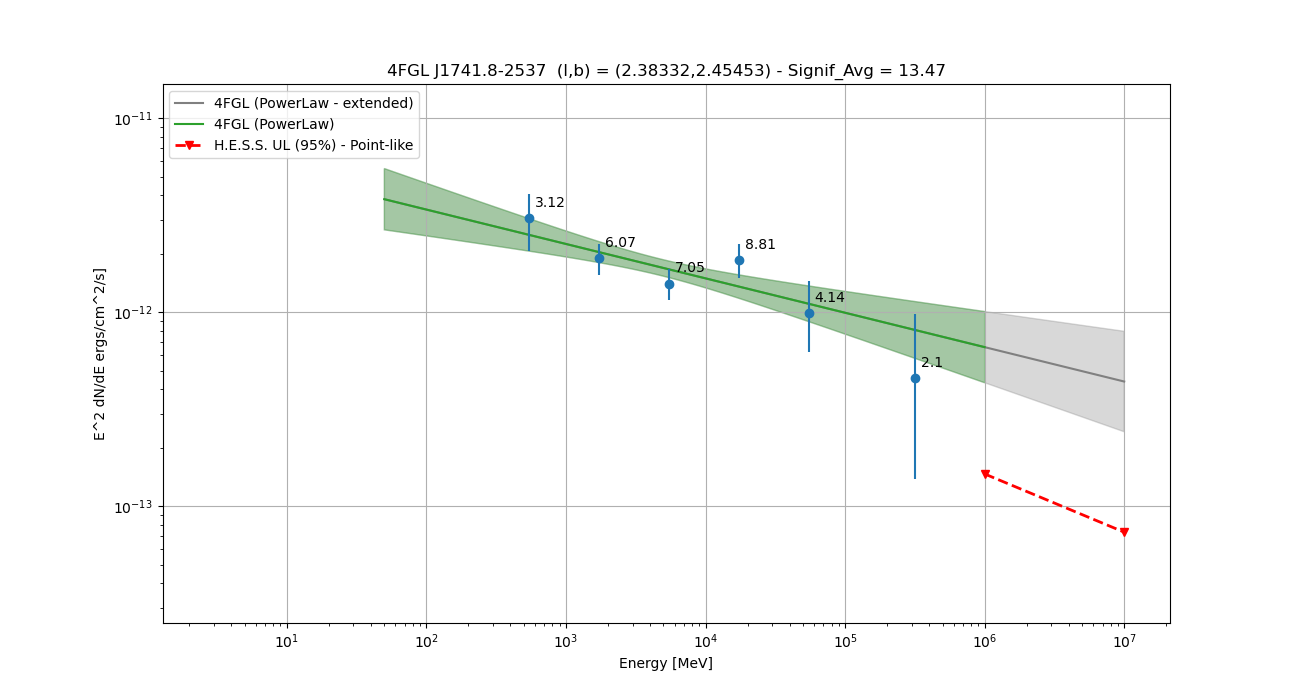}
        \includegraphics[width=0.49\textwidth]{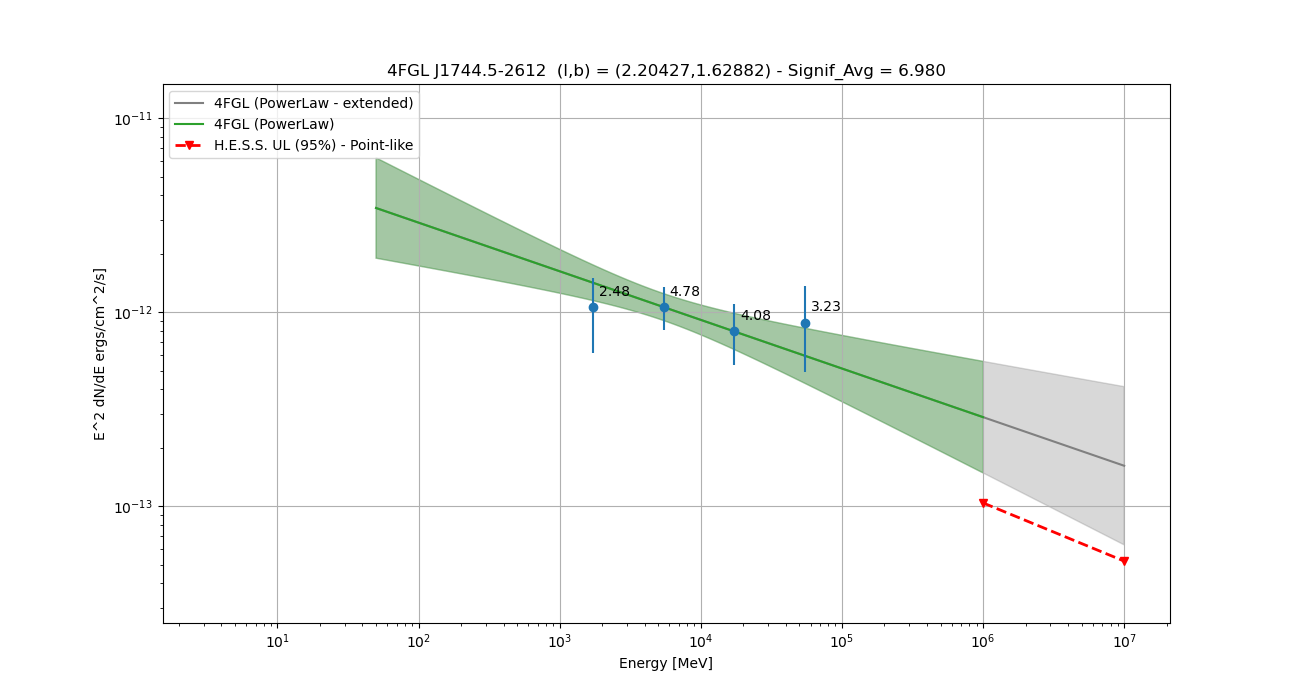}\\
        \includegraphics[width=0.49\textwidth]{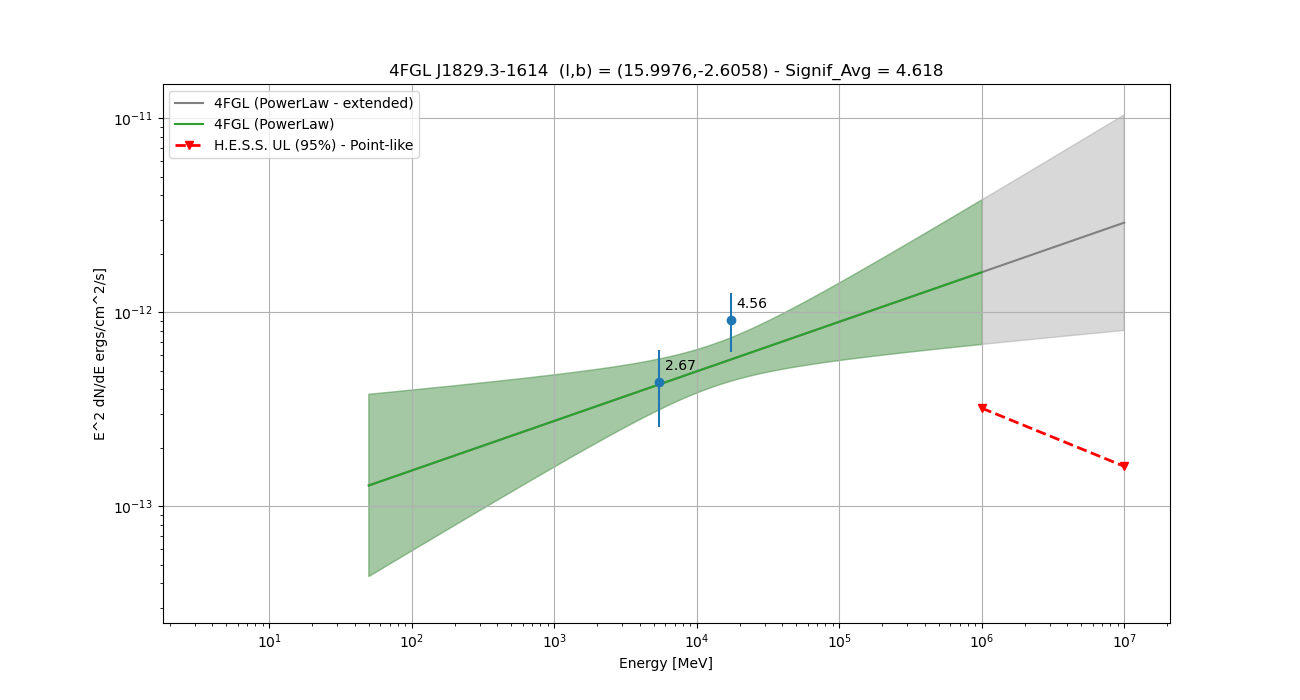}
        \includegraphics[width=0.49\textwidth]{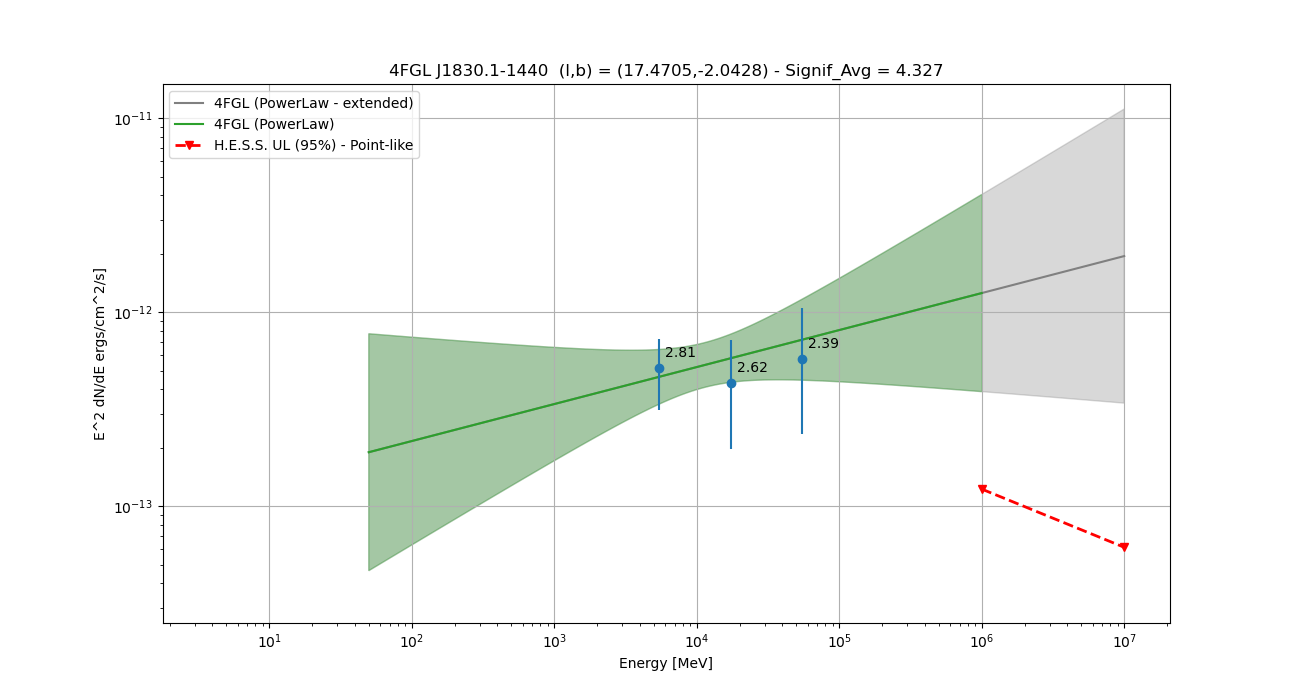}\\
        \includegraphics[width=0.49\textwidth]{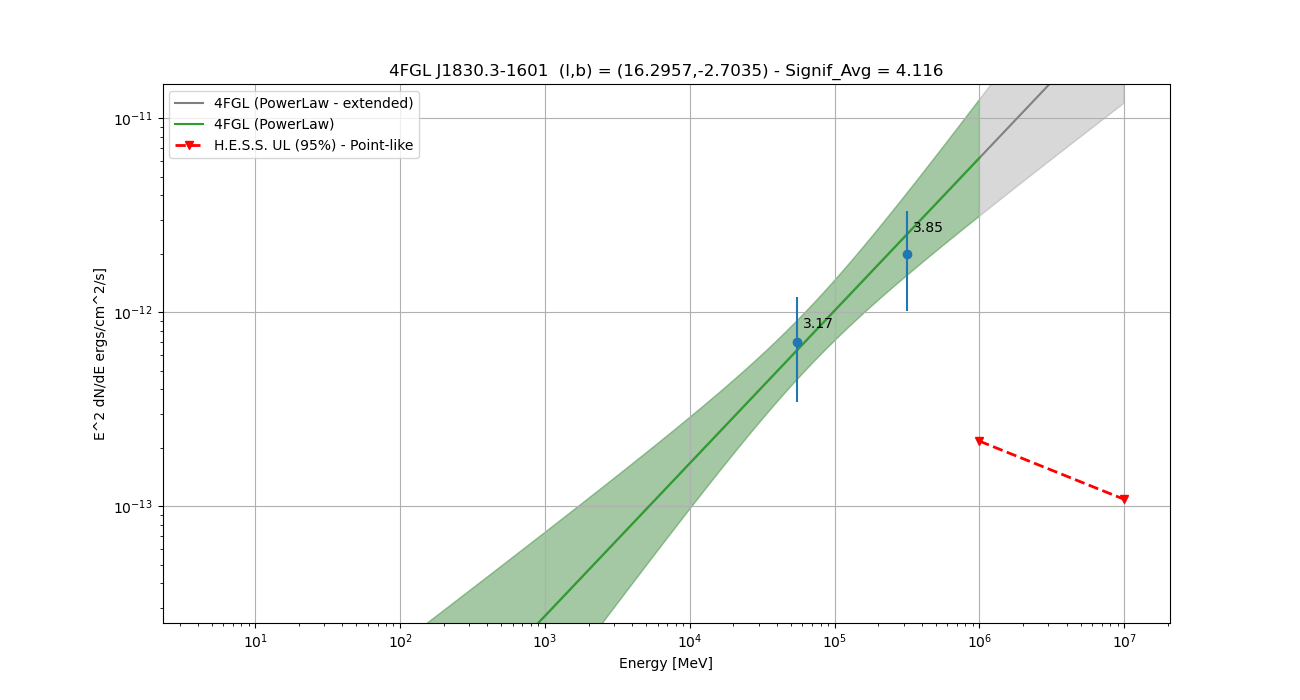}
    \caption{\textit{Continuation of Fig. \ref{fig_specs1}. }}
    \label{fig_specs2}
\end{figure*}

Four sources stand out as particularly constrained in our spectral comparison: 4FGL J1712.9-4105, 4FGL J1735.9-3342, 4FGL J1830.1-1440, 4FGL J1830.3-1601. As indicated in Table \ref{table_sources},  4FGL J1735.9-3342 and 4FGL J1830.3-1601 are coincident with the supernova remnants SNR G354.8-00.8 and SNR G016.2-02.7 respectively. The other sources have no identified associations. 
\

\begin{figure*}%[t!]
\centering
        \includegraphics[width=0.5\textwidth]{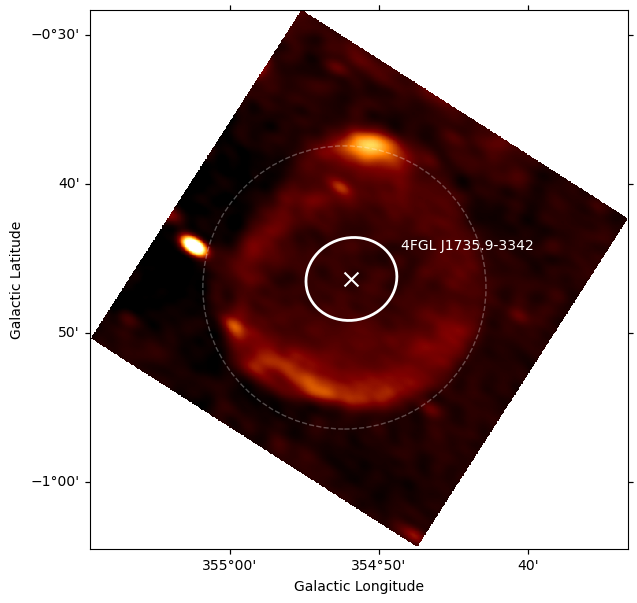}
    \caption{Radio map from the Molonglo Observatory Synthesis Telescope \cite{most} at 843 MHz towards the supernova remnant G354.8-00.8. The cross indicates the position of the \textit{Fermi}-LAT source 4FGL J1735.9-3342 and the ellipse represents the position uncertainty at 95\% Confidence Level. The dashed line indicates the boundary of the SNR as listed in \cite{greencat}.}
    \label{fig:fig_map_SNRG354.8-00.8}
\end{figure*}

The source coincident with the supernova remnant (SNR) G354.8-00.8 is localised in the center of the radio shell, as can be seen on Fig. \ref{fig:fig_map_SNRG354.8-00.8}. The \textit{Fermi}-LAT spectrum is hard, with a Power-Law spectral index of $1.67 \pm 0.16$. OH ($1720$~MHz) maser emission has been detected towards this source \cite{MaserOH}, indicating that the SNR shock is in interaction with a molecular cloud and favoring a hadronic origin for the gamma-ray emission. However the hard spectrum would favor a leptonic origin (through inverse Compton process), unless the emission can be explained by models such as described in \cite{gabici_rxj} with the presence of gas clumps in the vicinity of the SNR.
\

\begin{figure*}%[t!]
\centering
        \includegraphics[width=0.6\textwidth]{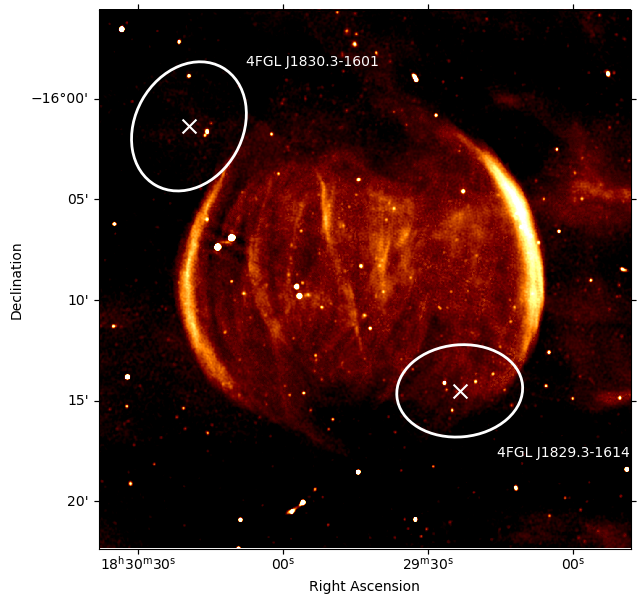}
    \caption{Radio map of the SNR G16.2-02.7 obtained by MeerKAT at
    1335.3 MHz \cite{meerkat}. 
    The crosses and ellipses shows the best-fit position and the 95\% CL position uncertainty of the \textit{Fermi}-LAT sources associated with this SNR.}
    \label{fig_map_SNRG16.2-02.7}
\end{figure*}

The case of SNR G016.2-02.7 is of particular interest as two of the selected sources (4FGL J1829.3-1614 and 4FGL J1830.3-1601) are coincident with this object. This remnant shows two lobes at radio wavelength \cite{SNRG16p2}, and each of the \textit{Fermi}-LAT source is indicated to be in spatial coincidence with one of the lobes. This is illustrated on Fig. \ref{fig_map_SNRG16.2-02.7}. In that case as well, the spectrum of the \textit{Fermi}-LAT source is hard, with a Power-Law spectral index of $1.21 \pm 0.22$. 
Assuming that the two sources are indeed associated to this SNR and assuming a distance of $6.5$ kpc and an age of $8$ kyr \cite{G16.2_meerkat}, the gamma-ray luminosity (between 50 MeV and 1 TeV) would be of $4\cdot10^{34}\mathrm{erg/s}$ and $3\cdot10^{34}\mathrm{erg/s}$ for 4FGL J1830.3-1601 and 4FGL J1829.3-1614, respectively.
In an hadronic scenario, assuming a characteristic time for $\pi^0$ production of $5\cdot10^{15}\cdot(1~\mathrm{cm^{-3}/n_H})$ seconds, the total energy in protons would be of at least $0.6\cdot10^{51}$ erg (for both sources), assuming a density at the shock of $0.6\mathrm{cm^{-3}}$, as inferred from the figures in \cite{G16.2_meerkat}. This is a very high fraction of the SN energy of $10^{51}$ erg.
In a leptonic scenario, the inverse Compton characteristic time is $7.6\cdot10^{14}$ seconds for $50$ GeV gamma-rays interacting on the CMB photon field, leading to a total energy in electrons of $5.3\cdot10^{49}$ erg. Assuming an electron to proton ratio of 1\% as in the Galaxy, the total energy in accelerated particles would again exceed the assumed SN energy of $10^{51}$ erg. 
A more detailed modeling of the source with more information from the environment would be needed to assess the validity of the association scenario.

\section{Summary and outlook}

Comparing the spectra of the \textit{Fermi}-LAT sources in the 4FGL-DR4 catalog with the flux ULs determined in the H.E.S.S. Galactic Plane Survey led to the identification of 13 constrained sources. The hard spectra observed in some of these \textit{Fermi}-LAT sources suggest efficient particle acceleration, possibly extending into the multi-TeV domain. However, the non-detection by H.E.S.S. places constraints on any TeV emission, implying the presence of a spectral break or cutoff, which may indicate aging accelerators, energy-dependent escape, or limited target material for hadronic interactions. Detailed modeling of the sources would help in understanding the acceleration mechanisms at play. 

Future gamma-ray observatories such as the Cherenkov Telescope Array Observatory (CTAO) will be key to resolving such cases \cite{ctabook}. With its superior angular resolution and sensitivity, CTAO will help to better characterize the sources in the GeV - TeV range. It will also enable detailed morphological and spectral studies of faint sources, probing spectral breaks and cutoffs beyond the current H.E.S.S. limits.

\bibliographystyle{JHEP}
\bibliography{biblio_hgps_fermi_icrc25}

\end{document}